# Probing the Continuum Limit in Non-Compact QED: New Results on Large Lattices*

M. Göckeler[a†], R. Horsley[a,b], V. Linke[c], P.E.L. Rakow[c],
G. Schierholz[b,a] and H. Stüben[d]

[a] Höchstleistungsrechenzentrum HLRZ,
c/o Forschungszentrum Jülich, D-52425 Jülich, Germany

[b] Deutsches Elektronen-Synchrotron DESY,
Notkestraße 85, D-22603 Hamburg, Germany

[c] Institut für Theoretische Physik, Freie Universität Berlin,
Arnimallee 14, D-14195 Berlin, Germany

[d] Konrad-Zuse-Zentrum für Informationstechnik Berlin,
Heilbronner Straße 10, D-10711 Berlin, Germany

**Abstract**

We present new Monte Carlo results in non-compact lattice QED with staggered fermions down to $m_0 = 0.005$. This extends our previous investigations on the nature of the continuum limit of QED.

*Talk presented by H. Stüben at The XII International Symposium on Lattice Field Theory "LATTICE 94", Bielefeld, Germany, Sept. 27 – Oct. 1, 1994.
†Supported by *Deutsche Forschungsgemeinschaft*.

# 1 Introduction

We report on the continuation of our investigation of lattice QED [1, 2]. The central question is whether the continuum limit of QED is trivial or not. We have been looking at non-compact QED with staggered fermions which has been found to have a second order chiral phase transition.

In [1, 2] we have demonstrated that Monte Carlo data are consistent with the triviality picture. Two statements could be made. Direct evidence for triviality is that the renormalised charge seems to vanish when the lattice spacing is sent to zero. Indirect support for triviality is that the chiral condensate $\sigma$ can be described by a mean field equation of state with logarithmic corrections.

The author of [3] independently comes to the same conclusions when analysing the available Monte Carlo data. Both statements were also arrived at by employing Schwinger-Dyson methods both analytically [4] and numerically [5].

Nevertheless in [6, 7] it was argued that the chiral condensate should obey a power law equation of state with non mean field critical exponents.

In order to try to resolve this controversy we have done new Monte Carlo runs on larger lattices down to a bare mass of $m_0 = 0.005$. Additionally we directly measured the so called susceptibility ratio $R$ [6]. It can be used to estimate the values of the critical coupling $\beta_c$ and the critical exponent $\delta$ without further assumptions.

Using meson mass ratios as approximations of $R$ the authors of [6] claim support for their *ansatz* of the equation of state. We have commented on the validity of these approximations in [2]. Here we will present direct measurements of $R$ and relate these results to findings from the equations of state.

# 2 Equations of State

Our *ansatz* for the equation of state motivated from [1] is

$$m_0 = A_1 \frac{\sigma^3}{\log^q(1/\sigma)} + A_2(\beta - \beta_c)\frac{\sigma}{\log^p(1/\sigma)} \qquad (1)$$

with parameters $\beta_c, p, q, A_1$ and $A_2$. In our previous fit [1] we fixed $q \equiv 1$ and obtained $\beta_c = 0.187(1)$ and $p = 0.61(2)$. The *ansatz* used in [6] reads

$$m_0 = A_\delta \sigma^\delta + A_b(\beta - \beta_c)\sigma^b \qquad (2)$$

with parameters $\beta_c, \delta, A_\delta, A_b$ and $b \equiv 1$ (hyperscaling). The published result is $\delta = 2.31$ with $\beta_c = 0.205$.

With all parameters free both equations are numerically similar. From (1) one can derive effective values for $\delta$ and $b$

$$\delta_{\text{eff}} = 3 + q/\log(1/\sigma), \quad b_{\text{eff}} = 1 + p/\log(1/\sigma). \qquad (3)$$

| mean field EOS + log. corrections (1) | | | |
|---|---|---|---|
| $\beta_c$ | $q$ | $p$ | $\chi^2/\text{dof}$ |
| 0.1887(3) | 0.55(6) | 0.53(2) | 5 |
| 0.1869(1) | $\equiv 1$ | 0.65(1) | 6 |
| power law EOS (2) | | | |
| $\beta_c$ | $\delta$ | $b$ | $\chi^2/\text{dof}$ |
| 0.1885(3) | 3.40(5) | 1.28(1) | 8 |
| 0.1954(2) | 2.60(2) | $\equiv 1$ | 30 |
| $\equiv 0.205$ | 2.03(1) | $\equiv 1$ | 64 |

Table 1: Fit results with MINUIT errors.

| chiral condensate at $m_0 = 0.01$ | | | |
|---|---|---|---|
| $L$ | $\beta = 0.19$ | $\beta = 0.20$ | $\beta = 0.21$ |
| 8 | 0.1700(24) | 0.1097(18) | 0.0788(15) |
| 10 | 0.1820(21) | 0.1244(18) | 0.0883(8) |
| 12 | 0.1835(13) | 0.1322(10) | 0.0917(6) |
| 16 | | 0.1344(9) | 0.0965(5) |
| $\infty$ | 0.1841(28) | 0.1348(13) | 0.0983(10) |

Table 2: Finite size dependence of $\sigma$. The $L = 10$ data are taken from the second paper of ref.[6].

In Table 1 we have collected results from fitting (1) and (2) to Monte Carlo data including our new points at $m_0 = 0.005$. We have done fits with all parameters free and also with fixed $q \equiv 1$ and $b \equiv 1$ respectively in order to compare with the results cited above. Values of $\chi^2$ per degree of freedom ($\chi^2/\text{dof}$) are also given. The best fit is shown in Figure 1.

Table 1 shows that our previous results [1] remain stable while the power law fit does not reproduce the parameters given in [6]. Using (2) with $b \equiv 1$ gives a relatively large $\chi^2/\text{dof}$. Fixing $\beta_c$ to a value of 0.205 (the monopole percolation threshold) increases $\chi^2/\text{dof}$ considerably.

Leaving all parameters free in (1) as well as in (2) leads to equivalent results. Positive values of $q$ and $p$ correspond to $\delta_{\text{eff}} > 3$ and $b_{\text{eff}} > 1$ as seen from (3). These results are also compatible with our previous results [1] but contradict the claims in [7, 8].

Finite size effects might in principle influence the results. For the chiral condensate we have checked that these effects are small on our largest lattices. They are of the order of the statistical errors for the data we used in our fits (see examples in Table 2).

To understand the finite volume dependence we have done calculations in the $\beta \to \infty$ limit. The leading dependence of the chiral condensate on the

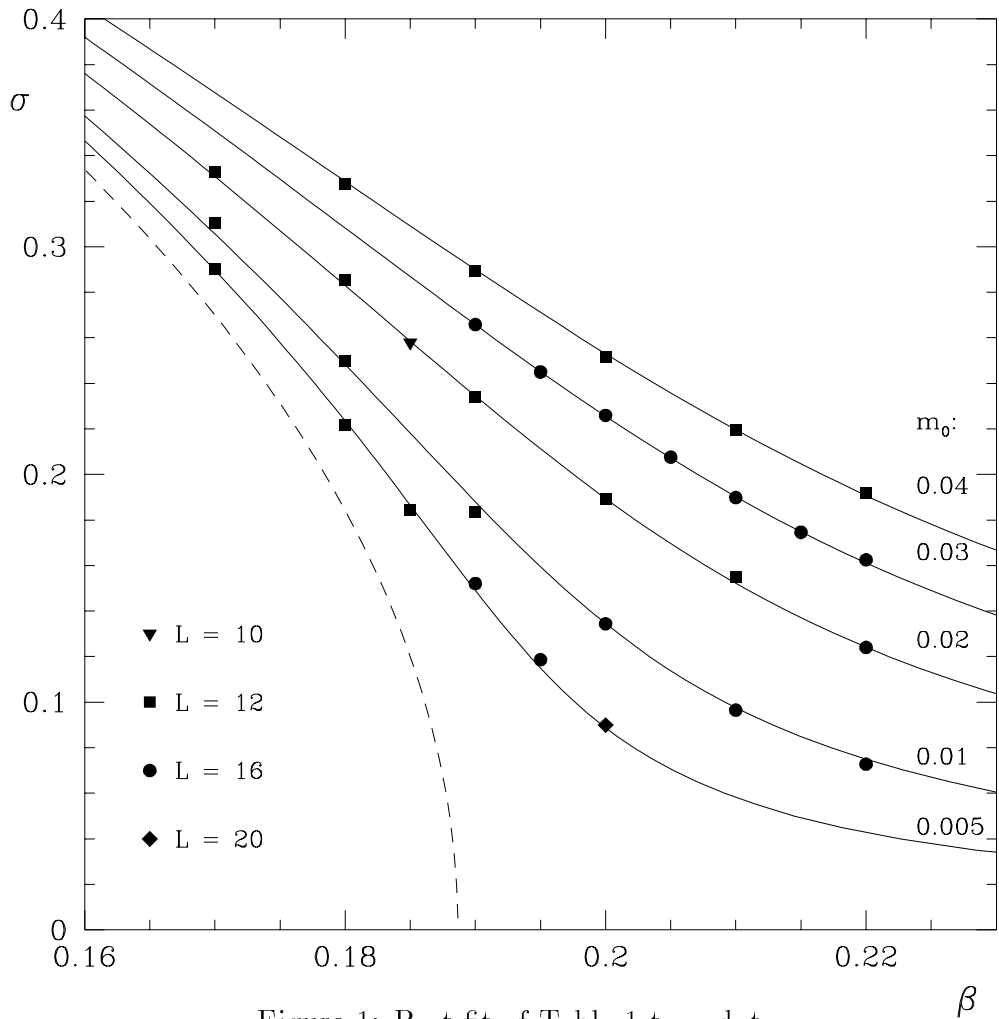

Figure 1: Best fit of Table 1 to $\sigma$ data.

linear lattice extension $L$ was found to be [2]

$$\sigma_L - \sigma_\infty \propto \sigma_L^3 \exp(-3.23\sigma_L L) + \ldots \qquad (4)$$

In Figure 2 we have plotted $\sigma_L$ against the r.h.s. of (4). One obtains reasonably straight lines if $\sigma_L \cdot L$ is not too small. If we use (4) to extrapolate to $L = \infty$ we obtain the open points. Some extrapolated values are given in Table 2. We have also done fits to the extrapolated data. These give results that are consistent with those of Table 1.

## 3   The Susceptibility Ratio

The susceptibility ratio $R$ is defined by

$$R := \frac{\partial \log \sigma}{\partial \log m_0} = \frac{\partial \sigma}{\partial m_0} \cdot \frac{m_0}{\sigma} \ . \qquad (5)$$

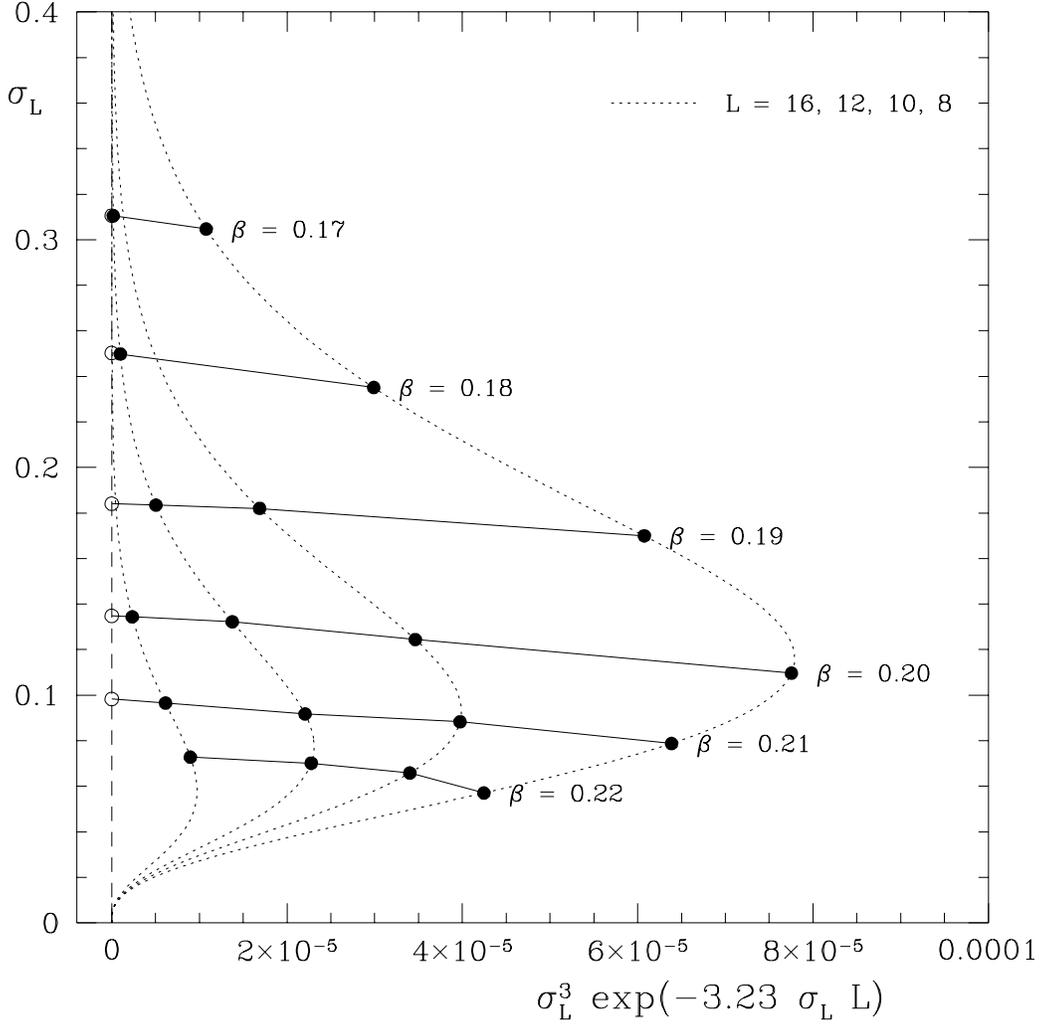

Figure 2: Finite size scaling of $\sigma$ at $m_0 = 0.01$.

To independently measure $R$ we employ the Ward identity $\partial\sigma/\partial m_0 = C_S^{\text{full}}(p = 0)$ where $C_S^{\text{full}}(p = 0)$ is the full scalar correlation function at zero momentum:

$$C_S^{\text{full}} = \frac{1}{V} \sum_{x,y} \left\{ \langle \bar{\chi}\chi(x)\bar{\chi}\chi(y) \rangle - \langle \bar{\chi}\chi(x) \rangle \langle \bar{\chi}\chi(y) \rangle \right\}.$$

The traces needed for the disconnected contributions were calculated with stochastic estimators. For each configuration we averaged over 20 vectors of random numbers.

The plot of $R$ versus $m_0$ can be used to estimate $\beta_c$ and $\delta$ [6]. In such a plot curves of constant $\beta > \beta_c$ bend upwards, curves of $\beta < \beta_c$ bend downwards and the curve for $\beta = \beta_c$ ends on the $y$-axis at $1/\delta$. In Figure 3 we show results of our measurements. From this plot one concludes that $\beta_c$ seems to lie near 0.19 and $\delta$ seems to be close to the mean field value 3.

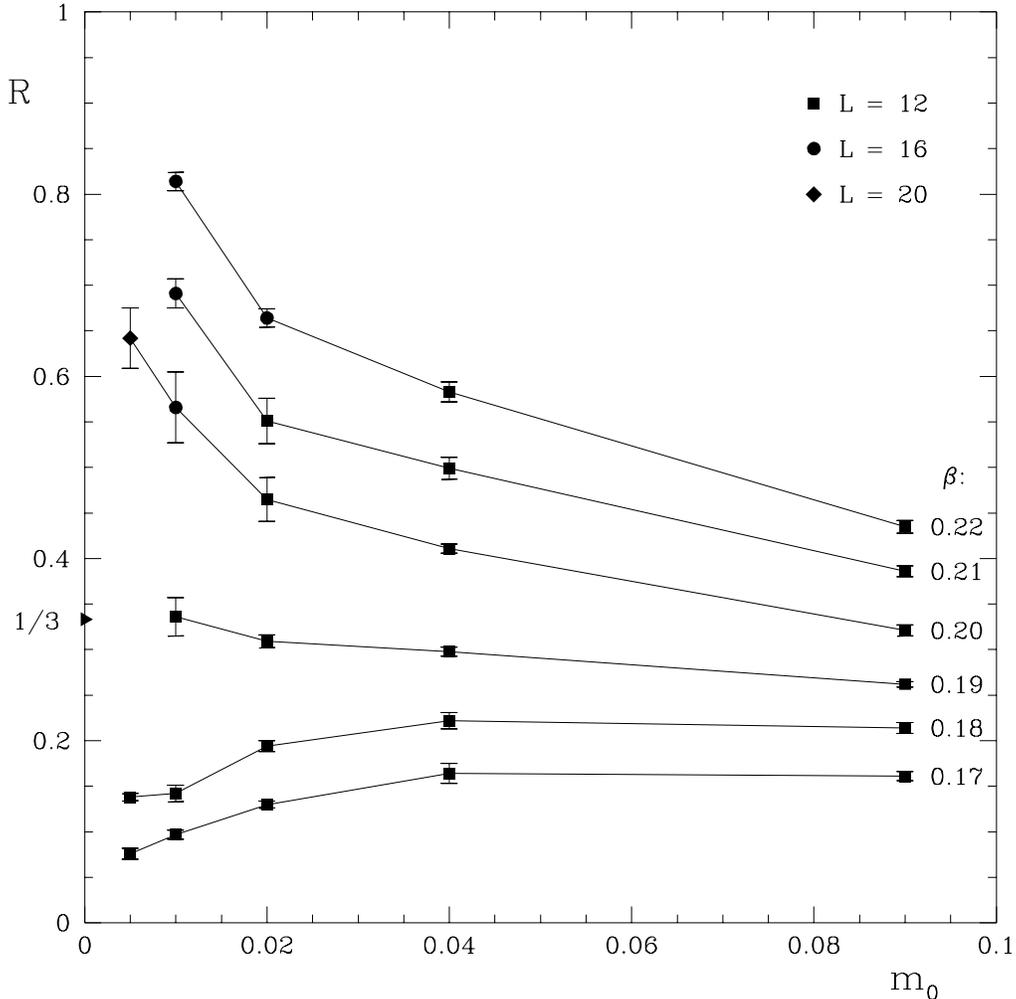

Figure 3: Measurements of $R$.

# 4 Conclusions

Our *ansatz* for the equation of state (1) describes the data well. If we also allow for hyperscaling violations in the power law *ansatz* (2) (i.e., keep $b$ as a free parameter) we get consistent results with our *ansatz*. In particular the values of $\beta_c$ agree and $\delta_{\text{eff}} > 3$. Demanding hyperscaling ($b \equiv 1$) does not give such a good fit. We have checked that finite size corrections to these results are small.

Our measurements of $R$ are consistent with the results from the three reasonable fits to $\sigma$. However, before we can draw definitive conclusions from the $R$ data a closer look at the critical region is needed. The measurements are being extended to $m_0 = 0.005$ and larger volumes.

The main conclusion is that at present all Monte Carlo data are consistent with the hypothesis that QED has a trivial continuum limit.

*Acknowledgement.* We thank ZIB Berlin and HLRZ Jülich for providing computer time.

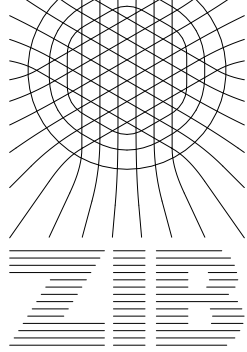

Meinulf Göckeler, Roger Horsley, Volkard Linke,
Paul Rakow, Gerrit Schierholz and Hinnerk Stüben

# Probing the Continuum Limit in Non-Compact QED: New Results on Large Lattices